# A High-Performance, Reconfigurable, Fully Integrated Time-Domain Reflectometry Architecture Using Digital I/Os

Zhenyu Xu, *Graduate Student Member, IEEE*, Thomas Mauldin, Zheyi Yao, Gerald Hefferman, and Tao Wei, *Member, IEEE*

*Abstract*—Time-domain reflectometry (TDR) is an established means of measuring impedance inhomogeneity of a variety of waveguides, providing critical data necessary to characterize and optimize the performance of high-bandwidth computational and communication systems. However, TDR systems with both the high spatial resolution (sub-cm) and voltage resolution (sub-$\mu$V) required to evaluate high-performance waveguides are physically large and often cost-prohibitive, severely limiting their utility as testing platforms and greatly limiting their use in characterizing and trouble-shooting fielded hardware. Consequently, there exists a growing technical need for an electronically simple, portable, and low-cost TDR technology. The receiver of a TDR system plays a key role in recording reflection waveforms; thus, such a receiver must have high analog bandwidth, high sampling rate, and high-voltage resolution. However, these requirements are difficult to meet using low-cost analog-to-digital converters (ADCs). This article describes a new TDR architecture, namely, jitter-based APC (JAPC), which obviates the need for external components based on an alternative concept, analog-to-probability conversion (APC) that was recently proposed. These results demonstrate that a fully reconfigurable and highly integrated TDR (iTDR) can be implemented on a field-programmable gate array (FPGA) chip without using any external circuit components. Empirical evaluation of the system was conducted using an HDMI cable as the device under test (DUT), and the resulting impedance inhomogeneity pattern (IIP) of the DUT was extracted with spatial and voltage resolutions of 5 cm and 80 $\mu V$, respectively. These results demonstrate the feasibility of using the prototypical JAPC-based iTDR for real-world waveguide characterization applications.

## I. Introduction

TIME-DOMAIN reflectometry (TDR) is an established means of measuring impedance discontinuities for a variety of waveguides [1]. TDR is a time-of-flight-based measurement technique, which measures the intensity of the reflections generated as a probe signal propagates along a waveguide. The resulting measurement of reflection intensity as a function of distance is termed the impedance inhomogeneity pattern (IIP) of the device under test (DUT). The key performance metrics of a TDR system are spatial resolution and voltage (alternatively termed vertical) resolution, which determines the location precision and the smallest measurable impedance variance achievable by the system. For a limited set of applications, such as soil moisture measurement, the resolution requirements are low [2]; consequently, efforts have been made to reduce the cost of TDR technology for these applications by reducing system performance [3]. However, the majority of TDR applications, such as locating faults in high-speed data links [4], [5], data bus authentication, and high-resolution distributed sensing [6], [7], require precision measurements. Consequently, TDR systems for PCB-level applications are typically stand-alone instruments of high complexity and cost used exclusively for product evaluation in the factory or laboratory rather than for fielded devices. A portable, low-cost TDR system capable of high-precision and high-resolution characterization fielded PCB-based devices, therefore, represents an important technology void.

Several steps can be taken to achieve the goal of building a low-cost TDR without comprising performance. If the TDR system can be integrated into the existing computational or communication platform and made reconfigurable, hardware overhead can then be significantly reduced. For example, a TDR system could be directly integrated into the I/O ports of a chip; using this approach, hardware resources can be used as general-purpose I/Os and reprogrammed for use as a TDR system when necessary, allowing online signal integrity testing and diagnosis. Importantly, a reduction in hardware overhead cannot result in compromised performance. Spatial resolution is determined by the analog bandwidth of both the transmitter (Tx) and receiver (Rx), as well as the sampling rate of Rx. The sensitivity is determined by the voltage resolution of the Rx. High Tx bandwidth, or high-speed probe signal in a modern digital platform, is relatively easy to achieve. In contrast, the voltage resolution and sampling rate of an analog-to-digital converter (ADC) are classic tradeoff pair due to jitter noise and analog bandwidth limitations [8], [9].

Manuscript received November 25, 2020; revised February 5, 2021; accepted February 11, 2021. Date of publication February 19, 2021; date of current version March 5, 2021. This work was supported in part by the National Science Foundation under Grant NSF-2027069 and Grant NSF-1439011. The Associate Editor coordinating the review process was Mohamed A. Abou-Khousa. *(Corresponding author: Zhenyu Xu.)*

Zhenyu Xu, Thomas Mauldin, Zheyi Yao, and Tao Wei are with the Department of Electrical, Computer, and Biomedical Engineering, The University of Rhode Island, Kingston, RI 02881 USA (e-mail: zhenyu_xu@uri.edu; thomas_mauldin@uri.edu; zheyi_yao@uri.edu; tao_wei@uri.edu).

Gerald Hefferman is with the Department of Electrical, Computer, and Biomedical Engineering, The University of Rhode Island, Kingston, RI 02881 USA, also with the Department of Radiology, Brigham and Women's Hospital, Boston, MA 02115 USA, and also with the Harvard Medical School, Boston, MA 02115 USA (e-mail: ghefferm@gmail.com).

Digital Object Identifier 10.1109/TIM.2021.3060586

Therefore, achieving high-voltage resolution and high sampling rates for real-time measurements is fundamentally challenging, as demonstrated by recent work by Orbach et al. [10], which sampled the reflection wave using multiple clock phases to increase the sampling rate, but is limited to 1-bit voltage resolution.

A possible solution to overcome this physical limitation is to sacrifice time efficiency. The technology described by Bishop et al. [11] applied equivalent time sampling (ETS) to achieve a high sampling rate using a 12-bit ADC, but an expensive 13-GHz track-and-hold chip was required due to the analog bandwidth limitation in order to match with the high ETS rate. An approach described by Trebbels et al. [3] also utilized the ETS method, while the analog bandwidth was increased by replacing the ADC with a 1-bit data converter (i.e., a comparator) and a digital-to-analog converter (DAC). The voltage comparisons, taken simultaneously in a typical ADC, are broken into different time slots to preserve resolution. These two methods reduce system cost in comparison with those standalone TDR instruments; however, additional active components (sample-latch and DAC) are required and, thus, do not represent a fully integrated and reconfigurable design.

A more recent approach described by Xu et al. [7] and Mauldin et al. [12] uses an analog-to-probability conversion (APC) concept to measure the signal in noisy environments using a differential digital I/O. However, this technique is limited by the offset voltage and hysteresis of the differential receiver/comparator [13]. In order to address this limitation, a quasi-triangle wave, namely, a probability density modulation (PDM) signal, is employed. This advantage of the quasi-triangle wave PDM signal is that it requires only passive additional hardware components; however, the system is still not fully reconfigurable since one of the pins of the differential receiver must be connected to the external passive hardware elements.

This article introduces a fundamentally different PDM scheme, namely, jitter-based APC (JAPC). JAPC capitalizes on the traditionally unwanted random jitter noise (RJN) of a clock signal. Together with a proposed noise reduction algorithm, JAPC fulfills the needs for APC in the previously reported iTDR design. Most importantly, RJN-based PDM completely obviates the need for any external circuit elements external to the IC, making the system fully reconfigurable. To the best of our knowledge, this represents the first demonstration of a high-performance, fully reconfigurable TDR system that does not require additional hardware. The specific TDR system described in this article was implemented using the digital I/O of a field-programmable gate array (FPGA) with modest use of logic resources and—most importantly—without PCB modification.

This article is organized as follows. Section II describes the design principle, specifically the proposed noise reduction algorithm and resolution analysis. Sections III details system implementation using a commercially available FPGA. Section IV describes the experimental setup and results. Section V includes discussion and concluding remarks.

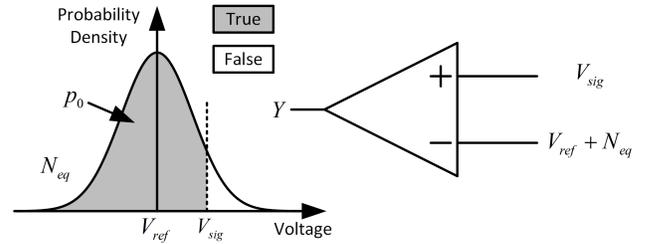

Fig. 1. Illustration of APC concept. $V_{\text{sig}}$: the instantaneous voltage of the input signal; $V_{\text{ref}}$: reference voltage applied to the inverting input; and $N_{\text{eq}}$: equivalent gaussian noise.

## II. WORKING PRINCIPLE

### A. Analog-to-Probability Conversion

APC is essentially a data converter that uses a comparator to compare an input signal against a noise signal and then uses the probability of "True" output to represent the voltage. The inverting input of the comparator is connected to the reference voltage, and the noninverting input is connected to the input signal. Assuming that both inputs are subject to Gaussian-like thermal noise [14], the probability of the "True" output ($Y$) can be written as

$$P(Y = 1, t) = P\{s(t) + N_p > V_{\text{ref}} + N_n\}$$
$$= P\{s(t) > V_{\text{ref}} + N_{\text{eq}}\} \quad (1)$$

where $s(t)$ is a clean signal and $N_p$ and $N_n$ represent noise at the noninverting and inverting inputs, respectively.

Since $N_n - N_p$ follows a Gaussian distribution, an equivalent noise $N_{\text{eq}}$, whose mean is 0 (ignoring the offset of the comparator) and the standard deviation is $\sigma_n^T$, is used to describe the noise of the system. If the noise distribution is stationary, given a certain input voltage $V_{\text{sig}} = s(t_0)$, the output probability is then a determined $p_0$, as shown in Fig. 1.

The voltage signal, $s(t)$, is changing over time, and its probability cannot be measured without repeated comparisons. APC leverages the fact that, given that the sampling time points are fixed with respect to the repeated reflection waveforms, the input voltage at all sampling points is a superposition of a constant signal (to measured) and random noise over the repeated measurements. Therefore, the APC naturally fits the ETS scheme, and the interactions between APC and ETS are further discussed in Section II-E.

Although the APC concept is straightforward, it faces several inherent barriers for real-world implementation. All commercial comparators have an offset voltage and hysteresis. The offset voltage results in the fact that, even when the reference voltage pin of the comparator is grounded, the actual $V_{\text{ref}}$ is not 0 [15]. This is an unavoidable effect, which results from the imperfectly matched input stage of the comparator. Most comparators have a mV-level offset voltage, leading to a significant offset probability and limited useful range. Hysteresis requires the input signal to have a large voltage difference with respect to the reference voltage in order to toggle the output, which equivalently reduces the noise and destroys the linearity. Therefore, Xu et al. [7] apply a triangle wave as a PDM signal so that the output of the comparator is always changing between 0 and 1 so that the hysteresis can be compensated. Section II-B describes a novel alternative PDM scheme to achieve the same result.

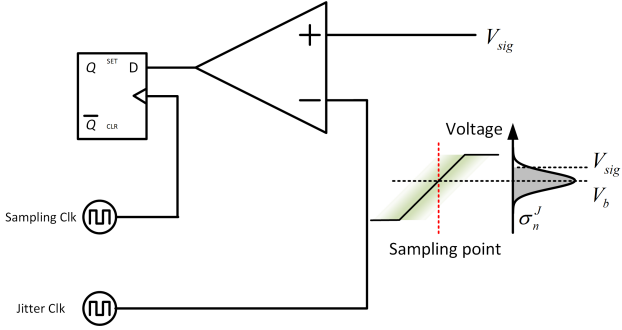

Fig. 2. AC equivalent schematic of the jitter-based PDM. $V_b$: the mean voltage applied to the inverting input at the sampling point.

### B. Jitter-Based PDM

PDM techniques compare the same input voltage against differing references, modulating the equivalent probability density function (pdf). As previously reported, a resistor–capacitor ($RC$)-based circuit can be used to generate pdf modulation signals [7], [12]. However, this approach requires the use of several circuit components outside the IC chip, adding necessary hardware to the system and disqualifying it as a fully reconfigurable iTDR.

The novel system described in this work alternatively solves this problem by exploiting the RJN from an internal clock. By definition, jitter is the deviation from the true periodicity of a presumably periodic signal. For a clock signal, the existence of RJN makes the actual arriving time of each rising edge fluctuate around an ideal arriving time (expected value), making the time of arrival a random process. Assuming that the expected arrival time is $t_0$ in each period, the actual arriving time, $t_n$, follows a Gaussian distribution as:

$$t_n \sim N(t_0, \sigma_j^2) \quad (2)$$

where the jitter noise, $\sigma_j$, is the standard deviation of the timing uncertainty.

The jitter noise is typically unwanted in any electronic system; however, in the proposed JAPC scheme shown in Fig. 2, an internal clock signal with an RJN ($\sigma_j$) is routed to the noninverting input of a comparator (negative input of a digital differential Rx). This internal clock is referred to as the jitter clock. By adjusting the phase relationship between the sampling clock and the jitter clock, it is possible to dynamically set the middle bias voltage, $V_b$, applied to the noninverting input of the comparator.

The slope (essentially the slew rate of the rising edge of a jitter clock) is denoted by $K$ with a unit of (V/s). Equivalently, this slope translates $\sigma_j$ into a voltage noise, denoted by $\sigma_n^J$, shown in Fig. 2, where $\sigma_n^J = K\sigma_j$. Importantly, $K$ and $\sigma_j$ are platform-dependent. The thermal noise, $\sigma_n^T$, also contributes to the total noise incident on the input buffer. Therefore, the total voltage noise, $N_{\text{total}}$, can be expressed as follows:

$$N_{\text{total}} \sim N\left(V_b, (\sigma_n^J)^2 + (\sigma_n^T)^2\right). \quad (3)$$

The jitter-based PDM solves the problems stemming from the use of comparators mentioned in Section II-A. The offset voltage can be compensated with $V_b$ by controlling the phase relationship between the jitter clock and the sampling clock. On the other hand, since $V_{\text{sig}}$ is around the mid-voltage of the comparator, when using the rising edge of the jitter clock signal for JAPC, the comparator's reference input always sweeps from low to high. In this case, its internal hysteresis behaves identical to its offset voltage, which can be overcome using the same approach as previously described.

In comparison to the previously reported PDM scheme in [7], jitter-based PDM is fully CMOS-integrated without the need for any external components, making it fully reconfigurable. It is worth mentioning that, since this design uses a digital input port, essentially a comparator, to measure reflection signals, the analog bandwidth always matches its counterpart, Tx (digital output port), in any platform. In addition, unlike conventional ADCs, comparators do not have a sample-and-hold function, making them broadband devices.

### C. Resolution Analysis

This section describes the performance metrics of the APC technique. All discussions in this section are based on two assumptions. First, the analog bandwidth is ultrawide, and no front-end filter is used; thus, thermal noise dominates. Second, the temperature of the PCB is stable, leading to a stable thermal noise level that does not vary over time. Under these two assumptions, the voltage resolution, $\Delta V$, archivable via APC is defined.

In the APC scheme, the probability is translated to voltage using the probability-to-voltage map (PVM). The average of the output of the comparator is an estimation of the expected probability. The accuracy of this estimation is a function of the total number of measurements. The estimation error, $E_p$, determines the probability resolution, $\Delta P$, which further translates to the voltage resolution, $\Delta V$, via the PVM. We will first define and derive probability error, $E_p$, and its corresponding resolution, $\Delta P$, in this section.

Given a specific input voltage, the expected probability $P(Y = 1)$ is a constant, namely, $p_0$. The output $Y$ follows a Bernoulli distribution. If an experiment is conducted $M$ times, and the results are named as $Y_i, i = 1.2, 3, \ldots, M$, the average, $\bar{Y}$, follows a binomial distribution. When $M$ is large enough [typically larger than $9(p_0/(1 - p_0))$ or $9((1 - p_0)/p_0)$] [16], the binomial distribution approximates a normal distribution, as shown in (4a)

$$\bar{Y} = \frac{1}{M} \sum_{i=1}^{M} Y_i \sim N\left(p_0, \frac{p_0(1 - p_0)}{M}\right) \quad (4a)$$

$$\bar{Y} \in \left[p_0 - 1.96\sqrt{\frac{p_0(1 - p_0)}{M}}, \; p_0 + 1.96\sqrt{\frac{p_0(1 - p_0)}{M}}\right] \quad (4b)$$

$$\Delta P = 3.92\sqrt{\frac{p_0(1 - p_0)}{M}}, \quad (\gamma = 0.95). \quad (4c)$$

Equation (4a) implies that $\bar{Y}$ equals $p_0$ when $M$ approaches infinity. The confidence level, $\gamma$, is defined in order to evaluate the performance. For example, if $\gamma = 0.95$, then it means that $\bar{Y}$ has a 95% probability that it locates within a range expressed in (4b). From another perspective, the APC has 95% confidence in distinguishing two probabilities with a difference of $\Delta P$. Probability uncertainty, $\Delta P$, is then defined as the width of this range, as expressed in (4c). For consistency, the probability uncertainty $\Delta P$ when $p_0 = 0.5$ is used to quantify the performance of APC in the following discussions. With $\Delta P$, the corresponding voltage resolution, $\Delta V$, can be calculated out using the PVM. It is worth noting that $\Delta V$ approaches 0 when $M$ approaches infinity, and in the APC scheme, the concept of quantization noise does not apply.

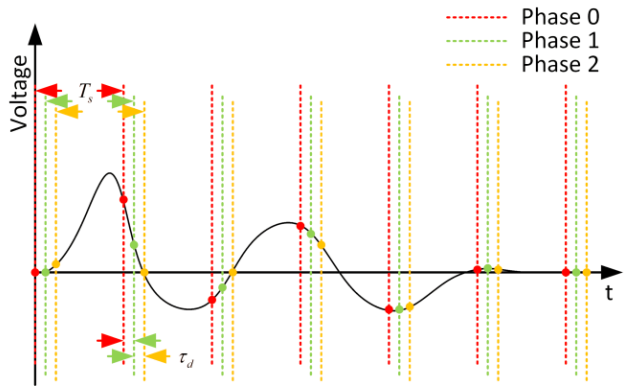

Fig. 3. Phase shift ETS. $T_s$: sample period of the sampling clock; $\tau_d$: the time delay between two phases.

### D. Phase Shift-Based ETS

ETS is an established means of equivalently increasing the sampling rate, and thus the spatial resolution, of TDR systems. Using ETS, multiple measurements of a repeating waveform with small phase differences are used to reconstruct the target waveform [11]. By shifting the phase relationship between the probe signal and sampling clock using internal phase-locked loops (PLLs), ETS can be employed in order to increase spatial resolution. The phase shift ETS first samples the waveform at a relatively low real-time sampling (RTS) frequency and then shift the phase of this sampling clock to resample the waveform. Thus, the interval between two consecutive samples in an RTS process is subdivided into many much smaller intervals, corresponding to a very high ETS rate. Fig. 3 illustrates the phase shift ETS scheme. The RTS rate is $1/T_s$, and the ETS sampling rate is $1/\tau_d$.

### E. Noise Analysis and Reduction

As a result of the high-voltage resolution of the system, iTDR measurement is sensitive to various noise sources, which reduces the effective voltage resolution from the theoretical value calculated in Section II-C. However, the iTDR utilizes both APC and ETS. These two schemes interact with each other and making the noise under different bands mixed with each other and can hardly be removed by conventional filters. A series of noise reduction techniques are implemented along with the JPAC topology in order to reduce noise and maximize voltage resolution.

In order to better explain these noise reduction techniques, we will first identify and categorize the dominant noise sources. Both RTS- and ETS-based techniques are used in an iTDR to measure a waveform. Accordingly, noise sources can be categorized based on the sampling domain (RTS or ETS domain) that they belong to.

To complete a waveform measurement, iTDR follows a three-step process.

*1) Step 1:* Transmit the probe signal and perform RTS with a time interval of $T_s$ (sampling rate of $1/T_s$) for $P$ samples (0 through $P - 1$). It, therefore, takes a time duration of $P \times T_s$ to complete Step 1.

*2) Step 2:* Repeat Step 1 $M$ times (0 through $M - 1$), and take an average of the $M$ measurements at each RTS time point ($P$ time points in total) to resolve measurements $S(0), S(T_s), \ldots, S((P - 1)T_s)$, resulting in a total of $P$ samples. Thus, it takes a time duration of $P \times M \times T_s$ to complete Step 2.

*3) Step 3:* Shift the phase between probe signal and sampling time point by $\tau_d$, where $\tau_d = T/J$. Repeat Step 2 to measure $S(\tau_d), S(T + \tau_d), \ldots, S((P - 1)T_s + \tau_d)$, a total of $P$ samples. Since the time duration needed for a PLL to perform a phase shift is small (12 clocks cycles, or $12T_s$), we can ignore it when $M$ is relatively large. This process is repeated until the RTS time interval, $T_s$, is filled with ETS time interval, $\tau_d$. As a result, within each RTS time interval, $J$ samples are obtained. These $J$ samples are referred to as a SET. For example, in the first RTS time interval, $J$ samples are $S(0), S(\tau_d), \ldots, S((J - 1)\tau_d)$. Together, they form the first SET, namely, SET 0. In total, a TDR waveform consists of $P$ SETs (SET 0 through SET $P-1$). Equivalently, the total number of samples is $P \times J$. Thus, it takes a time duration of $P \times M \times J \times T_s$ to complete the whole process (Step 3).

Fig. 4 illustrates this measurement process based on a simple example, where $P$ is set to 3 for easy demonstration. The $Y$-axis represents the RTS domain, and the $X$-axis represents the ETS domain (reconstructed waveform). Steps 1–3 are also signposted accordingly in the RTS domain on Fig. 4. The repetitions of Step 1 in Step 2 and the third to the last Step 3 are omitted.

The first noise to consider is system tones. In a digital system, all flip-flops are switching synchronized to the system clock, draining the current from the power regularly. Therefore, the voltage on a power rail usually has a periodic ripple with the same frequency as the system clock. Inevitably, this supply voltage ripple affects all other components inside the IC, such as the reference voltages of the comparator. Therefore, system tones are naturally synchronized to the ETS, where the sampling clock is shifting according to the main system clock. Consequently, the system tones are unavoidably captured in the TDR waveform and represent a source of the ETS-domain noise. Fortunately, this type of noise is highly repeatable and can be considered as an add-on noise to the original TDR waveform. Thus, it can be easily removed by subtracting the background waveform (i.e., making a measurement without transmitting a probe signal) from the measurement waveform.

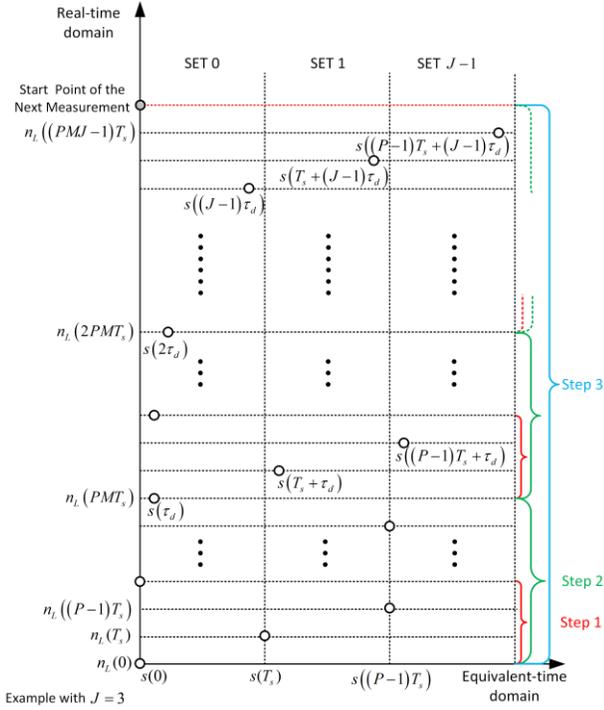

Fig. 4. Influence of low-frequency noise in ETS and APC. $M$: average time of APC; $s(t)$: synchronized waveforms; $n_L(t)$: low frequency noise; $P$: the number of real-time samples in each TDR probing, which is set to 3 as an example; $M$: average times; $J$: the entire number of ETS phases; $T_s$: sampling period; and $\tau_d$: minimum phase delay.

Since APC is based on probability measurement, it must be assured that all experiments (samples) are independent. Therefore, the noise in the RTS domain is subdivided into two general categories: high-frequency noise and low-frequency noise with respect to the real-time sampling rate. In general, a bandlimited noise with a cutoff frequency of $f_c$ has its autocorrelation function falls to 0 at $1/(4f_c)$. High-frequency noise has large bandwidth, and thereby, the first cross-zero point of its autocorrelation function is closer to 0, which means that such noise losses its correlation quickly in the time domain. High-frequency noise mainly stems from thermal and jitter noise, which typically has a bandwidth of over 1 GHz. Correspondingly, the experiments (samples) taken shall be independent if the sampling rate is smaller than 4 GHz. Because of the independency, these noises are attenuated by the averaging step in the JAPC measurements. Thus, no special technique is needed to mitigate high-frequency noise.

In contrast, low-frequency noise varies slowly with respect to the sampling rate and mainly stems from vibrations and slow changes in temperature. Such noise could stay at a similar absolute voltage for more than $1 \mu s$, making all samples made inside this time period not independent. Unlike system tones in the ETS domain and high-frequency noise in the RTS domain, the low-frequency noise is difficult to remove. A further complicating fact is that the use of ETS results in low-frequency noise appearing as high-frequency noises in final TDR waveforms; as a result, standard filtering techniques cannot be easily applied. A special noise mitigation technique is, therefore, needed, which is presented in the following.

As shown in Fig. 4, the low-frequency noise, denoted as $n_L(t)$, at different time points are marked on the vertical axis (RTS domain) through a complete measurement process. The low-frequency noise at $j_{th}$ phase shifted position in SET $p$, namely, $N_L^p(j)$, can be written as follows:

$$N_L^p(j) = \frac{1}{M} \sum_{m=0}^{M-1} n_L(\text{PMjT}_s + \text{PmT}_s + pT_s). \quad (5)$$

The low-frequency noise sources are slow, time-varying functions; as a result, the low-frequency noise in Step 1, where $P$ samples are taken at an RTS rate of $1/T_s$ (typically, in the range of a few hundred MHz), is considered as a constant.

Therefore, the low-frequency noises at $j_{th}$ phase shifted position in all SETs (SET 0 through SET $P-1$) are identical, namely, $N_L(j)$. Thus, (5) can be reduced to the following equation:

$$N_L(j) = \frac{1}{M} \sum_{m=0}^{M-1} n_L(\text{PMjT}_s + \text{PmT}_s). \quad (6)$$

Equation (6) shows that $N_L(j)$ is a constant among all SETs, suggesting that the low-frequency noise can be removed as long as one $N_L(j)$ waveform, i.e., $N_L(j)$ as a function of phase-shifted positions in one SET, is known. This waveform can be used as the noise waveform reference. We can subtract this noise waveform reference from all SETs to eliminate the low-frequency noise. Hence, intentionally leaving a set that contains no reflection signal as the noise waveform reference serves as a means of low-frequency noise reduction. This technique has the additional benefit of also reducing system tone since the system tones are periodic functions on TDR waveform with a period of $T_s$.

Notice that multiple SETs exist because $P$ samples are made in a single probing cycle. The primary purpose is to increase the time efficiency of the iTDR (it takes $P^2 \times M \times J \times T_s$ to record the same length if only one sample is made in each probing cycle). However, it turns out that it also enables the proposed noise reduction algorithm.

## III. HARDWARE IMPLEMENTATION

### A. Configuration of a Bidirectional Differential IO (BiDi-Diff-IO) on an FPGA

In order to empirically evaluate the system, a fully reconfigurable iTDR enabled by JAPC was implemented on an FPGA (Xilinx ZCU104). Fig. 5 shows the exemplary schematic of the proposed iTDR built around a bidirectional differential IO port (BiDi-diff-IO) where the output tristate buffer (OBUFT_N) on the negative side of the BiDi-diff-IO serves as the transmitter of the proposed iTDR, which transmits the probe signal to the DUT. The data input of this output tristate buffer (OBUFT_N) is always set to logic 1. For the majority of the time, the OBUFT_N is detached/disabled by configuring its output to have high impedance by setting its tristate control port (T_N) as low. As a result, during this time, there exists no transmitter in the iTDR system. In order to generate a probe signal (ideally an impulse function), the tristate control port (T_N) of OBUFT_N is driven by an impulse signal provided by logic resources in the FPGA fabric. Upon the arrival of this impulse logic signal, OBUFT_N is attached/enabled for a very short period of time. The impulse-like probe signal for TDR measurement is transmitted to the DUT through PAD_N. PAD_N is internally connected with both the output of OBUFT_N and the inverting port (PAD_N) of the comparator, i.e., the differential input. The reflected signal from the DUT is collected by PAD_N. The noninverting port (PAD_P) of the BiDi-diff-IO is used to collect jitter-based PDM signal, or jitter clock, for JAPC measurement. The jitter clock signal is generated by the internal positive output (OBUT_P) of this BiDi-diff-I/O. The internal jitter clock connects to the input port of OBUT_P, and the tristate control port (D_P) is always enabled.

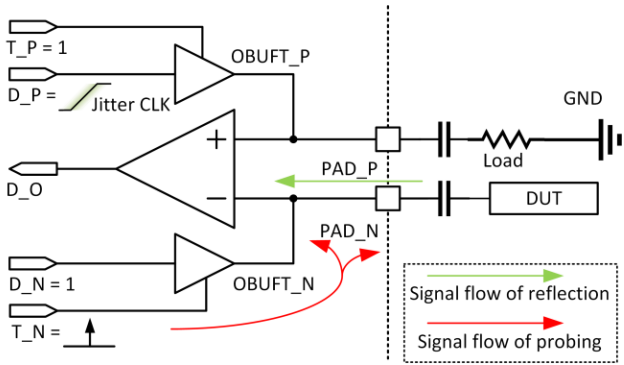

Fig. 5. Schematic of the TDR based on the bidirectional differential buffer. OBUFT_P (N): three-state buffer; T_P (N): three-state buffer enable signal; the buffer drives the output when it is "1" in this case; D_P (N): output data for the three-state buffer; D_O: output of the comparator; PAD_P (N): physical PAD that connected to the printed circuit board; and Load: the match resistor that exists at the receiver.

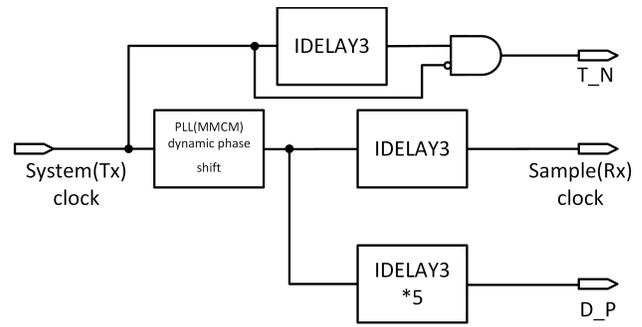

Fig. 6. Clock structure of the TDR. PLL: phase-locked loop; T_N: control pin of the OBUFT_N; and D_P: the signal to drive the OUBFT_P.

It is worth noting that, during the transmission time (the duration of the impulse-like function), the inverting input of the comparator (PAD_N) is forced to a logic high voltage, which is larger than the jitter clock signal, applied on the noninverting port (PAD_P). Therefore, the output of the comparator is always 0 during the launching time, and the reflection signal is blocked. In other words, a "blind spot" exists. Hence, it is advantageous to keep the duration of this impulse-like function as short as possible, in order to achieve the smallest "blind spot."

It is also worth noting that this configuration is built around the BiDi-diff-IO of an FPGA. Theoretically, either the trace connected to the PAD_P or PAD_N could be the DUT depending on the setup. The only limitation is that the two traces of differential pair cannot be measured simultaneously. In addition, the traces must be ac-coupled, which is almost always the case in high-speed data links, to maintain proper biasing.

The positive trace of this differential pair connects PAD_P with an external load. Depending on PCB design, this load can be a matched termination or simply an open circuit. In this experimental implementation, both were investigated with no observed influence on the measurement results.

### B. Clock Structure

The clock structure of the system is shown in Fig. 6. In this configuration, the impulse-like function (probe signal) connected to the tristate control port of OBUT_N is derived from the system clock. By "AND ing" a delayed version of the system clock with an inversed version of the system clock, we are able to generate an impulse-like function with a duration as small as 1 ns. A controllable logic-based delay unit is employed to generate the delayed version of the system clock.

A PLL with dynamic phase shift capability is used to realize the phase shift between the launch time of probe signal and sampling time to fulfill the needs required by Step 3 (ETS) in the measurement process. Xilinx Ultrascale+ FPGA is capable of generating an 11.12-ps phase shift, or an 89.6-GHz ETS rate at most, which is not the limitation compared with the analog bandwidth of the digital IOs (typically 1.6 GHz).

The jitter clock signal connected to the D_P is derived from the sampling clock. A set of controllable logic-based delay components is used to align the midpoint of the rising edge of the jitter clock signal to the sampling clock. In addition, these delay components enlarge the jitter of the jitter clock with respect to the sampling clock. Each delay component provides a maximum of 1.1 ns of delay with 512 taps, which is not sufficient for clock alignment for most cases. Therefore, five units are cascaded in the design. Limited by the platform, all the five delay lines must share the same delay if they are connected in series, which decreases the phase adjustment resolution. Therefore, a sixth independent delay unit is applied to the sampling clock path, allowing fine alignment adjustments.

## IV. EXPERIMENTAL RESULTS

### A. Experimental Setup

In order to demonstrate and evaluate the described design, an FPGA mezzanine card (FMC) was designed and fabricated. As shown in Fig. 7, the FMC board is connected to the SMA connectors of the differential pairs through ac-coupled capacitors. The initial system configuration only utilizes the negative output of the differential pair, while the positive output is terminated by a load so that it functions as a single-port iTDR. When the X2 port is not used for TDR, it can be reconfigured as a single-end IO port. The second configuration is a symmetrical differential pair, which can be configured as a differential IO port when the pair is not used for TDR. The remaining board components were not utilized. An adapter board was fabricated in order to connect standard interfaces to the SMA connectors.

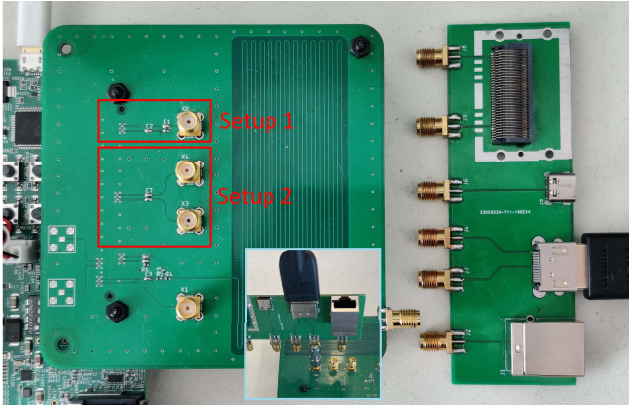

Fig. 7. Photograph of the experimental setup. The motherboard (on the left-hand side) is the Xilinx ZCU104 Evaluation board; the daughterboard (PCB in the middle) is an FMC board to lead out the differential pairs; and the adaptor board (PCB on the right-hand side) connects the interfaces to SMA connectors.

In this experiment, an HDMI cable was used as the DUT and tested under four different configurations: 1) no cable (open connector); 2) a connected HDMI cable without receiver (open end); 3) a connected HDMI cable with the receiver (terminated); and 4) a curved cable.

An autocalibration process was first conducted in which the phase difference between the jitter clock and the sampling clock is aligned to make the output probability approach 0.5 via adjusting the delay of the delay units on both the sampling clock path and the jitter clock path, as shown in Fig. 6. Thus, the midpoint of the rising edge of the jitter clock signal is almost aligned with the sampling clock, and the offset voltage of the comparator has been compensated. Since the rising edge of the jitter clock is quasi-linear around its midpoint, the voltage of the reflection signal is linearly proportional with a probability of the output of the differential input, or the comparator, of the BiDi-diff-IO. Therefore, we directly utilize probability to determine the waveform.

The experiments were conducted using following parameters: the RTS interval, $T_s$, was 10 ns, i.e., the RTS rate was 100 MSPS (mega samples per second); the length of the TDR waveform, $PT_s$, was set to 100 ns, where $P = 10$ (10 SETS); and the ETS interval, $\tau_d$, was 17.8 ps, i.e., the ETS rate was 56 GSPS (giga samples per second). Therefore, there are 560 phase-shifted positions ($J = 560$) in each SET and a total of 5.600 ETS (PJ) samples in a TDR waveform. The probe signal, or the impulse, was 600 mV in amplitude and 1 ns in duration. For all experiments, $M$ was set to 1000, which approximately provided a vertical resolution (probability) of 0.006 with $\gamma = 0.95$, corresponding to a voltage resolution of 80 $\mu$V. It is worth noting that these measurement parameters can be easily reconfigured on this SoC/FPGA chip. For comparison purposes, the above-mentioned parameters were held constant for all measurement results in the following. The jitter noise ($\sigma_n^J$) was measured as 8 mV, and the thermal noise ($\sigma_n^T$) is less than 1 mV as the resistances are all 50 $\Omega$. Therefore, the jitter noise dominates. Experimental noise reduction evaluation.

The effectiveness of the proposed noise reduction techniques was experimentally evaluated. Fig. 8(a) shows the measured waveforms with and without a probe signal. It is clear that the system tones are prominent in both waveforms. By subtracting the waveform without the probe signal from the waveform with probe signal (system tone reduction technique), the compensated TDR waveform (blue trace) was generated, as shown in Fig. 8(b), and a significant noise reduction was observed.

Fig. 8(b) also shows the TDR waveform (red trace) after applying the low-frequency noise reduction technique. The data in SET0, a total of 560 samples, were used as the reference noise waveform to be subtracted from the other nine SETs. The noise was drastically reduced, showing the effectiveness of the proposed low-frequency noise reduction technique.

Also, a quick comparison between two techniques in Fig. 8(b) clearly shows that the low-frequency noise reduction technique is superior to the system tone reduction technique. This is due to the fact that the low-frequency noise reduction technique also removes system tones. The low-frequency noise reduction technique was then applied to all TDR waveforms reported in the following.

It is worth noting that the noise reduction technique can effectively decrease the required number ($M$) of repeated RTS measurement in Step 2 (APC) without compromising voltage resolution. Thus, total measurement time can be substantially reduced.

*B. Experimental TDR Evaluation*

Fig. 8(c)–(f) shows the plots of the results of all four experiments, including: 1) no cable (open connector); 2) a connected HDMI cable without a receiver (open end); 3) a connected HDMI cable with a receiver (terminated); and 4) a terminated but curved cable.

Fig. 8(c) shows the TDR waveform without an HDMI cable. The first peak at around 10 ns is a measure of the impulse-like probe signal when it is transmitted to the DUT via PAD_N. The negative impulse-like probe signal is so strong that it saturates the measurement. A strong reflection peak at around 12.9 ns was observed at the HDMI connector. This is a strong reflection from an open circuit, which also saturates the measurement. Echo signals were observed following the strong reflection. According to the measured TDR waveform, it takes 2.9ns for the probe signal to make the round trip from the I/O port on the FPGA chip to the HDMI connector and back. Typically, the wave speed in a PCB trace is about $1.5 \times 10^8$ m/s. A 21.75 cm trace was measured to exist between the I/O port and the HDMI port, which agrees with the physical setup.

Fig. 8(d) shows the result when a 193−cm HDMI cable is connected to the HDMI connector, while the receiver end is powered OFF (open circuit). In this case, a reflection is expected at the HDMI connector because of the inevitable impedance mismatch at the connection. A large additional reflection is also expected to be generated at the end of the second HDMI cable. The experimental result matches the expectation, and the time difference between the two reflections was 19.46 ns. Considering that the speed of light in HDMI cable is $1.85 \times 10^8$ m/s, the measured time difference agrees well with the physical length of the cable setup. Echo signals were observed following the strong reflection in the measured TDR, as shown in Fig. 8(d).

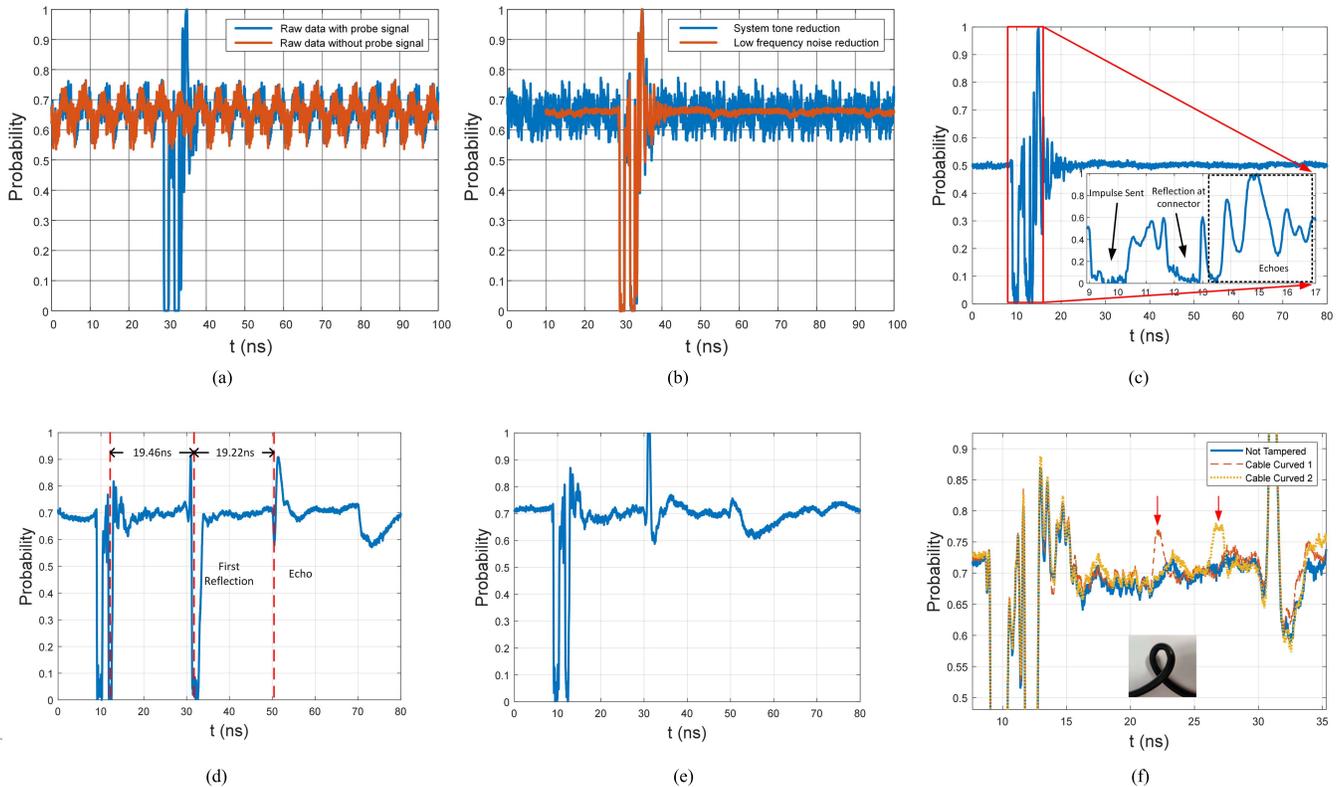

Fig. 8. Time-domain reflection measured with the proposed TDR. (a) Raw data and system tones. (b) Comparison between two noise reduction algorithms. (c) Reflection waveform measured when there is no HDMI connected. (d) Reflection when an HDMI cable is connected with the load opened. (e) Reflection when an HDMI cable is connected with the load matched. (f) Reflection when the HDMI cable is bent at two different locations.

Fig. 8(e) shows the result when a 193−cm HDMI cable is connected to the HDMI connector, and while the receiver is powered ON (matched termination). A small negative reflection was observed at the termination of the HDMI cable, again the consequence of expected impendence mismatch at the cable connections. Though the reflection still saturates the measurement range, it is not a large reflection as the measurement range is naturally small, and the width is much smaller than other reflections.

The fourth experiment was designed to evaluate whether the proposed iTDR can measure small reflections caused by bending the HDMI cable. In this test, the HDMI cable is bent by about 120° at two different locations, as shown in Fig. 8(f), which also illustrates the resultant reflections along the HDMI cable, illustrating that the system can detect even small impedance inhomogeneities.

## V. Conclusion and Discussion

This article describes a novel technique, JAPC, which obviates the need for external components in order to make high-resolution TDR measurements. Experimental testing demonstrates that a fully reconfigurable and highly integrated TDR (iTDR) can be successfully implemented using an FPGA chip without using any external circuit components. An HDMI cable was employed as the DUT to empirically evaluate the feasibility of using the system for real-world applications. The IIP of this DUT was successfully extracted using the proposed JAPC-iTDR. The spatial resolution is determined by the bandwidth of TX, which was approximately 5 cm under the testing conditions described above. The voltage resolution is approximately $80\mu$V when $M$ is 100,000. According to the FPGA usage report, the JAPC-iTDR uses one PLL (MMCM), 1.945 LUTs, and 1.594 registers, which are the numbers that can likely be further reduced with further optimization efforts.

Since our ultimate purpose is to achieve a fully reconfigurable and iTDR without any modification on hardware and external components so that such a TDR can be implemented on any I/O ports of an FPGA when needed, our design cannot do self-calibration, which means that the absolute voltage is not obtainable in most cases and its resolution not comparable with commercial TDRs or VNAs. However, we do not consider them as drawbacks because they are designed for different applications in the first place.

To conclude, JAPC-iTDR is a fully reconfigurable, low overhead, and high-performance TDR system with the potential to allow for the accurate characterization and optimization of fielded high-performance computing and communication infrastructure.

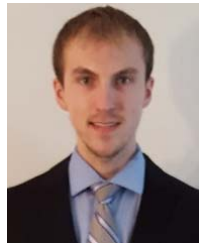

**Thomas Mauldin** received the B.S. and M.S. degrees in electrical engineering from The University of Rhode Island, Kingston, RI, USA, in 2018 and 2019, respectively, where he is currently pursuing the Ph.D. degree with the Electrical, Computer, and Biomedical Engineering Department.

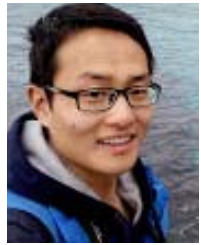

**Zheyi Yao** received the B.S. and M.S. degrees in electrical engineering from the Nanjing University of Science and Technology, Nanjing, Jiangsu, China, in 2014 and 2017, respectively. He is currently pursuing the Ph.D. degree with the Electrical, Computer, and Biomedical Engineering Department, The University of Rhode Island, Kingston, RI, USA.

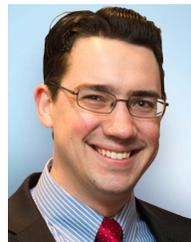

**Gerald Hefferman** received the B.I.D. degree in industrial design from the Pratt Institute, Brooklyn, NY, USA, in 2009, the Diploma degree in pre-medical studies from the Harvard Extension School, Cambridge, MA, USA, in 2013, and the M.D. degree with a scholarly concentration in medical technology, innovation, and entrepreneurship from the Warren Alpert Medical School, Brown University, Providence, RI, USA, in 2019.

He is currently a Radiology Resident Physician with the Brigham and Women's Hospital, Boston, MA, USA, a Clinical Fellow in Radiology with the Harvard Medical School, Boston, MA, USA, and an Adjunct Professor of engineering with The University of Rhode Island, Kingston, RI, USA.

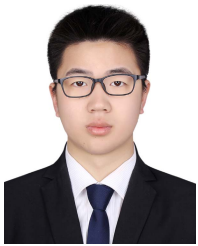

**Zhenyu Xu** (Graduate Student Member, IEEE) received the B.S. degree in electrical engineering from Nanjing Tech University, Nanjing, Jiangsu, China, in 2018. He is currently pursuing the Ph.D. degree with the Electrical, Computer, and Biomedical Engineering Department, The University of Rhode Island, Kingston, RI, USA.

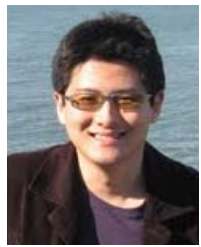

**Tao Wei** (Member, IEEE) received the B.S. degree from Nanjing Tech University, Nanjing, Jiangsu, China, in 2006, and the M.S. and Ph.D. degrees from the Missouri University of Science and Technology, Rolla, MO, USA, in 2008 and 2011, respectively.

He is currently an Associate Professor of electrical engineering with The University of Rhode Island, Kingston, RI, USA. His research interests include optoelectronic devices, embedded systems, and sensing systems.